\documentclass[10pt]{iopart}
\usepackage{graphicx}

\begin{document}

\title[Edge mode velocities in QHE]{Edge mode velocities in the quantum Hall effect from a dc measurement}

\author{P. T. Zucker and D. E. Feldman}

\address{Department of Physics, Brown University, Providence, Rhode Island 02912, USA}
\ead{dmitri\_feldman@brown.edu}
\vspace{10pt}
%\begin{indented}
%\item[]February 2014
%\end{indented}

\begin{abstract}
Because of the bulk gap, low energy physics in the quantum Hall effect is confined to the edges of the 2D electron liquid. The velocities of edge modes are key parameters of edge physics. 
They were determined in several quantum Hall systems from time-resolved measurements and high-frequency ac transport. We propose a way to extract edge velocities from dc transport in a point contact geometry defined by narrow gates. The width of the gates assumes two different sizes at small and large distances from the point contact. The Coulomb interaction across the gates depends on the gate width and affects the conductance of the contact. The conductance exhibits two different temperature dependencies at high and low temperatures. The transition between the two regimes is determined by the edge velocity. An interesting feature of the low-temperature $I-V$ curve is current oscillations as a function of the voltage. The oscillations emerge due to charge reflection from the interface of the regions defined by the narrow and wide sections of the gates.
 
\end{abstract}

% Uncomment for PACS numbers
%
% Uncomment for keywords
\vspace{2pc}
 
% Uncomment for Submitted to journal title message
\submitto{\NJP}
%
% Uncomment if a separate title page is required
%\maketitle
% 
% For two-column output uncomment the next line and choose [10pt] rather than [12pt] in the \documentclass declaration
%\ioptwocol
%

\section{Introduction}

One of the central ideas of condensed matter physics is universality. Despite the great variety and complexity of materials properties, it is frequently possible to make precise quantitative predictions from a few features of a system, such as its symmetry or topological order. Universality often comes hand in hand with scaling and is exhibited at a low energy scale. A textbook example is quantum critical behavior.  On the other hand, insulators do not have low-energy excitations in the bulk and thus their low-energy physics is trivial. In an ordinary insulator this would be the end of the story. Topological insulators \cite{ti1,ti2} differ by the presence of topologically protected gapless edge and surface states, such as a helical Fermi liquid on the surface of a 3D topological insulator or Luttinger liquids on the edges of 2D $Z_2$ insulators and quantum Hall systems. Moreover, the principle of bulk-edge correspondence establishes a deep connection between the universal aspects of edge physics and the nature of the bulk excitations \cite{wen}.

The research on edge transport in the quantum Hall effect (QHE) has long been motivated by both the intrinsic interest of edge physics and its use as a tool to understand the properties of the bulk. In particular, most experimental methods that have been implemented or proposed to understand the puzzle \cite{rmp52} of the QHE at the filling factor $5/2$ involve edge properties
% \cite{dolev,radu,lin,yf13,baer,li,overbosch09,seidel,wang10a,bid,fdt1,viola,fdt2,fdt3,fradkin98,dassarma05,11,12,14,grosfeld06,hou,15,sternreview,willett09,bishara09,hrwy,fp331,willett10,16,kang,yang15,yang09,gervais,barlas}
\cite{dolev}-\cite{barlas}. 
These methods include interedge tunneling 
%\cite{dolev,radu,lin,yf13,baer}
\cite{dolev}-\cite{baer}, probing upstream neutral modes
%\cite{li,overbosch09,seidel,wang10a,bid,fdt1,viola,fdt2,fdt3}
\cite{li}-\cite{fdt3}, and various 
interferometry schemes
%\cite{fradkin98,dassarma05,11,12,14,grosfeld06,hou,15,sternreview,willett09,bishara09,hrwy,fp331,willett10,16,kang,yang15}
\cite{fradkin98}-\cite{yang15}. 
The latter approach stimulated much work on interferometry at other filling factors which brought several enigmatic experimental results \cite{goldman05,marcus,moty}. In order to understand those results and interpret the experiments about topological orders in the second Landau level, it is imperative to achieve a better understanding of edge physics in the QHE.

One of the basic pieces of data about an edge is the velocity of charge propagation. Conceptually, the most straightforward approach to its measurement is based on a time-resolved experiment: A voltage pulse is applied at the source and the edge velocity is deduced from the time it takes for the Hall voltage to appear at the probe 
%\cite{ashoori,zhitenev,ernst,sukhodub,kamata,kumada}
\cite{ashoori}-\cite{kumada}.
Other methods involve the resonant transmission or absorption of microwaves  \cite{talyanskii,andreev}
and ac transport measurements at GHz frequencies in nanostructures \cite{gabelli,hashinaka,G13}. To the best of our knowledge only one approach that does not rely on high frequency measurements has been implemented: The velocity is extracted from the voltage dependence of the current transmitted through a quantum Hall interferometer \cite{interf-exp,interf}. Such an approach was successful in the integer QHE  \cite{interf-exp}. Its extension to the fractional QHE is difficult \cite{goldman06} because of challenges in the experimental implementation and theoretical interpretation of interferometry in the fractional regime. In this paper we propose a different way to measure the edge velocity. Our method uses dc transport through a single quantum point contact (QPC). It requires a simple modification of several recent tunneling experiments 
\cite{dolev}-\cite{baer}
and its implementation will likely be easier than that of interferometry.

Our setup is illustrated in Fig. 1. A 2D electron gas is depleted under metallic gates on the two sides of a QPC. This defines two QHE edges around the upper and lower gates. A QHE liquid exists between the edges. The width of the gates depends on the distance from the QPC and assumes two different sizes at short and long distances. A voltage bias $V$ is applied between the upper and lower edges. The bias drives a tunneling current $I$ through the QPC. 
This geometry can be used for a test \cite{yf13} of the theory of the $5/2$ state, proposed in Ref. \cite{yf14}. We focus on the weak tunneling regime. The edge velocity can be extracted from the temperature dependence of the zero-bias conductance $G=dI/dV\big|_{V=0}$.

The simplest theoretical description of edge transport, the chiral Luttinger model \cite{wen}, predicts a power law for the zero-bias conductance as a function of the temperature:

\begin{equation}
\label{1}
G\sim T^{2g-2},
\end{equation}
where $g$ is a universal exponent that depends on the topological order. The prefactor in front of the temperature is nonuniversal and depends on the transmission coefficient of the QPC. Experiment does not agree with the predicted universality. For some filling factors, such as $5/2$ and $8/3$, the observed $g$ agrees reasonably well with some of the theoretical proposals for the topological order \cite{yf14,baer}.  At other filling factors, such as $1/3$ and $7/3$, the experimental results for $g$ are considerably higher than its universal theoretical value \cite{baer,roddaro}. The discrepancy of the experimental and theoretical $g$ was explained by the effect of the Coulomb interaction between portions of the edge on the different sides of the gate \cite{papa04,papa05,yf13}. Such an interaction is strong as long as the gate is not much wider than its distance from the 2D electron gas and the width of the edge.

One may see the effect of interaction on $g$ as an unfortunate obstacle to probing topological orders with Eq. (\ref{1}). For the measurement of the edge velocity, however, such an effect is a boon. Indeed, the point contact can only feel what happens within the thermal length $l_T\sim \hbar v/T$, where $v$ is the speed of the edge mode. If the voltage bias $V$ is greater than the temperature then the thermal length is substituted with $l_V\sim \hbar v/e^*V$, where $e^*$ is the charge of the current carriers that tunnel through the QPC. Thus, the observed $g$ is determined by the interaction across the narrow part of the gates as long as $l_T\ll a$, where $a$ is the length of the narrow part. If $l_T\gg a$ then the observed $g$ is determined by the interaction across the wide part of the gates. The crossover temperature between the regimes with two different $g$'s depends on the edge velocity $v$.

To go beyond an order of magnitude estimate of the velocity one needs to compare the experimental temperature dependence of the conductance with the theory. Such a theory is given below.

Besides the Coulomb effect, the observed $g$ may be affected by edge reconstruction \cite{cw}.
In edge reconstruction, pairs of contrapropagating modes emerge along the edge. This may change $g$ in the high-temperature regime but does not affect the tunneling exponent at low temperatures. Indeed, at sufficiently large length scales, disorder localizes such mode pairs. Hence, their existence becomes unimportant, if $l_T$ exceeds the localization length.
This length depends on the sample details. For example, the Weizmann group did not observe edge reconstruction in the integer QHE and saw edge reconstruction in the fractional QHE only at the lengths of the order of 
microns \cite{edge-rec}. On the other hand, the Harvard group observed a significant edge reconstruction effect \cite{rec-yacoby} at the filling factor $\nu=1$ at the length scale of 20 $\mu$m. In our calculations we assume that there is no edge reconstruction at the relevant length scales. It is possible to extend our theory to an edge-reconstructed system but this goes beyond the scope of the present work. Several ways exist to probe the presence or absence of edge reconstruction in a given system. This includes, for example, local thermometry, developed in Ref. \cite{rec-yacoby}.

This  paper has the following structure. In the next section we derive a general expression for the tunneling current. Its limiting cases are analyzed in the Appendix. We discuss how to use the tunneling current to determine the velocity in Section 3. Section 4 summarizes our results. 

Everywhere, we set $\hbar=k_B=1$.

\section{Tunneling current}

\subsection{Model}

\begin{figure}
\centering
\includegraphics[width=2.5in]{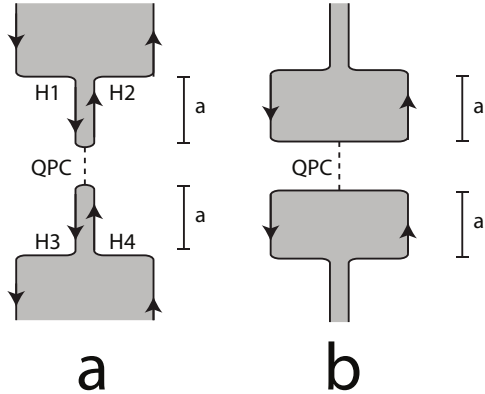}
\caption{Charge propagates around the shaded gates in the direction of the arrows. Dashed lines show tunneling at the QPC. a) Interaction across the gates is stronger near the QPC. b) Interaction across the gates is 
completely screened near the QPC.}
\end{figure}

We consider the two geometries, illustrated in Fig. 1. In both cases the interaction across the gates assumes two different values at short and long distances from the point contact. 
The interaction strength changes at the distance $a$ from the QPC. In Fig. 1a, the interaction is stronger at short distances. In Fig. 1b the interaction is weaker at short distances. 
For simplicity, we assume that the interaction across the gates is absent in their wide section due to electrostatic screening. Thus, we assume that the width of the wide part of the gates considerably exceeds the distance between the gates and the 2D electron gas as well as the width of the QHE edge. The latter is set by the electrostatics of the sample as discussed in Ref. \cite{csg}.
It is straightforward to generalize our calculations to the case of nonzero interaction everywhere.

Coulomb interaction may be present between the edges on the different sides of the QPC. We will neglect that interaction. Thus, we assume that the width of the gates is much less than $a$. For the same reason we disregard the horizontal portions of the edges, labeled H1, H2, H3 and H4 in Fig. 1. This justifies our assumption that the interaction abruptly changes at the distance $a$ from the QPC.

We will limit our discussion to the simplest integer and fractional QHE states at $\nu=1$ and $1/3$. Its extension to other filling factors is straightforward. 
We introduce the coordinates on the two edges so that the tip of the contact is located at $x=0$ on both edges. $x$ measures the length along the edges. It increases in the charge propagation direction   as shown by the arrows in Fig. 1.
The action assumes the form

\begin{equation}
\label{2}
S=\int dt dx [L_u+L_l] -\int dt\hat T,
%\nonumber\\
\end{equation}
where $L_{u,l}$ are the Lagrangian densities for the upper and lower edges and $\hat T$ is the operator describing charge tunneling between the edges. The Lagrangians of the upper and lower edges \cite{wen} are expressed in terms of Bose fields 
$\phi_i$, $i=u,l$:

\begin{equation}
\label{3}
L_i(x,t)=-\frac{1}{4\pi}[\partial_t\phi_i(x,t)\partial_x\phi_i(x,t)+u(x)\partial_x\phi_i(x,t)\partial_y\phi_i(y,t)\big|_{y=-x}+v(\partial_x\phi_i(x,t))^2],
\end{equation}
where the physical meaning of the fields $\phi_i$ comes from their relation to the charge densities $\rho_i=\sqrt{\nu}e\partial_x\phi_i/2\pi$
and the electric currents $j_i=-\sqrt{\nu}e\partial_t\phi_i/2\pi$, $v$ is the speed of the edge mode in the absence of the interaction across the gates, and $u$ describes the strength of that interaction.
We assume contact interaction between the nearest points $x$ and $-x$ on the opposite sides of a symmetric gate. Such a choice of the interaction corresponds to an effective low-energy long-scale model. Fig. 1a shows a setup with $u(x)=0$ at $|x|>a$, $u(x)={\rm const}$ at $|x|<a$. In Fig. 1b, $u(x)=0$ at $|x|<a$ and $u(x)={\rm const}$ at $|x|>a$. Since the interaction is repulsive, $u(x)\ge 0$.
The tunneling operator

\begin{equation}
\label{4}
\hat T=\Gamma\exp(i\sqrt{\nu}[\phi_u(x=0)-\phi_l(x=0)])+H.c.,
\end{equation}
where $H.c.$ stays for Hermitian conjugation,
describes the transfer of a single charge $e^*=\nu e$ across the QPC. The tunneling operator also contains contributions that describe transfer of higher charges. They are less relevant at low energies and will be neglected below.
We assume that the tunneling amplitude $\Gamma$ is small enough to justify the use of perturbation theory. As will be clear from the calculations below, this means that

\begin{equation}
\label{5}
\Gamma < E_c^\nu[{\rm max}( e^*V,k_B T)]^{1-\nu},
\end{equation}
where $E_c$ is the relevant ultraviolet cutoff energy of the order of the energy gap.
Our model is related to the model of a quantum wire with a weak impurity from Ref. \cite{tr}.

\subsection{Tunneling}

To describe the voltage bias, we follow the steps, outlined in Refs. \cite{fg}. We assume that no tunneling between the edges was possible in the distant past, $t=-\infty$. Then the tunneling is turned on. Assuming that this happened long before $t=0$, the details of the turning on procedure do not matter for the current at $t=0$. Since in the absence of the tunneling, the charges $Q_u$ and $Q_l$ of the upper and lower edges conserve, the voltage bias at $t=-\infty$ can be understood as the difference of the chemical potentials $\mu_u-\mu_l=eV$ between the upper and lower edges which are assumed to be in thermal equilibrium at $t=-\infty$. The difference of the chemical potentials at 
$t=-\infty$ can be set to zero in an appropriate interaction representation \cite{fg} as discussed below. We first add to the Hamiltonians of the upper and lower edges the contributions $-\mu_iQ_i=-\int dx \rho_i(x)\mu_i=-\sqrt{\nu}e\mu_i\int dx \partial_x\phi_i/2\pi$, $i=u,l$, where $Q_{u,l}$ are the total charges of the upper and lower edges. Simultaneously, we set both chemical potentials to zero and introduce a time-dependence into the operators which do not commute with $Q_u$ or $Q_l$. In particular, Eq. (\ref{4}) becomes

\begin{equation}
\label{6}
\hat T(t)=\Gamma\exp(-ie\nu V t)\exp(i\sqrt{\nu}[\phi_u(x=0)-\phi_l(x=0)])+H.c.
\end{equation}
Finally, we shift the fields $\phi_i$ by time-independent corrections such that the edge Lagrangians assume the old form (\ref{3}) up to an irrelevant constant.

Next, we need to identify the tunneling current operator. We define it as the time derivative of the electric charge of the lower edge.

\begin{eqnarray}
\label{7}
I=\dot{Q}_{l}=-\dot{Q}_{u}=i[H,Q_{l}]=i[\hat T,Q_l] \nonumber\\
=-ie^*\Gamma\exp(-ie\nu V t)\exp(i\sqrt{\nu}[\phi_u(x=0)-\phi_l(x=0)])+H.c.
\end{eqnarray}
The electric current at $t=0$ is given by the average $\bar I =\langle S(-\infty,0)I S(0,-\infty)\rangle$, where the angular brackets denote the average with respect to the initial equilibrium density matrix. $S(t_1,t_2)$ is the evolution operator. We expand the latter to the first order in  the tunneling amplitude $\Gamma$:

\begin{equation}
\label{8}
\bar I=\langle i\int_{-\infty}^0 dt [\hat T(t), I(0)] \rangle.
\end{equation}
The above expression can be evaluated with the observation that the right hand side is a combination of the average exponents of free fields $\phi_{u,l}$. Free fields satisfy the identity 

$$
\langle \exp(A)\exp(B)\rangle = \exp(\frac{1}{2}\langle A^2\rangle+\langle AB \rangle + \frac{1}{2}\langle B^2 \rangle ).
$$
Thus, one finds

\begin{equation}
\label{9}
\bar I=e^*|\Gamma|^2\int_{-\infty}^\infty dt (e^{ie^*Vt}-e^{-ie^*Vt})\exp(2\nu\langle \phi_u(x=0,t)\phi_u(0,0)- \phi^2_u(0,0)\rangle),
\end{equation}
where we use the fact that the correlation functions are the same for $\phi_u$ and $\phi_l$.

\subsection{Correlation function}

The next step is to compute the correlation function $G(t)=\langle \phi_u (0,t)\phi_u(0,0)-\phi^2_u(0,0)\rangle$. In what follows we omit the index $u$. 

Consider first Fig. 1a, where $u(x)=0$ at $|x|>a$ and $u(x)=u_0$ at $|x|<a$.
We start with the equation of motion for the field $\phi$ in the absence of the tunneling $\hat T$:

\begin{equation}
\label{10}
\partial_{xt}\phi(x)+v\partial_{xx}\phi(x)-\partial_{x}[u(x)\partial_x\phi(-x)]=0.
\end{equation}
This is equivalent to

\begin{equation}
\label{11}
\partial_{t}\phi(x)+v\partial_{x}\phi(x)-u(x)\partial_{x}\phi(-x)=C(t),
\end{equation}
where $C(t)$ is a coordinate-independent operator. To identify it we rely on chirality of the Lagrangian (\ref{3}) at $x<-a$. Chirality implies that the solution $\phi(x<-a,t)$ must be exactly the same as in the absence of the interaction $u(x)=u_0$ upstream (for a detailed discussion of the relation of chirality and causality, see Refs. \cite{fdt1,fdt2,fdt3}). The latter solution is well known \cite{vdo}:

\begin{equation}
\label{12}
\phi_0(x,t)=\frac{2\pi\rho x}{\sqrt{\nu}e}+\sqrt{\frac{2\pi}{L}}\sum_{q>0} \frac{1}{\sqrt{q}}(e^{iq(x-vt)}b_q+e^{-iq(x-vt)}b^\dagger_q),
\end{equation}
where $\rho$ is the average charge density, $L\rightarrow\infty$ is the total length of the edge and $b^\dagger_q,b_q$ are boson creation and annihilation operators with the standard commutation relations 
$[b_q,b^\dagger_k]=\delta_{q,k}$. We see that $C(t)=2\pi v \rho/\sqrt{\nu}e$ is independent of both coordinate and time. One can remove $C(t)$ from Eq. (\ref{11}) by subtracting a time-independent function 
of the coordinate from $\phi$. This is equivalent to the convention that $\rho=0$. Thus, we have to solve the equation of motion

\begin{equation}
\label{13}
\partial_{t}\phi(x)+v\partial_{x}\phi(x)-u(x)\partial_{x}\phi(-x)=0.
\end{equation}

With $\rho=0$, the solution at $x<-a$ can be cast in the form $\phi(x,t)=\phi_0(x/v-t)$. What about $x>-a$? A direct substitution verifies the following general solution at $-a<x<a$:

\begin{equation}
\label{14}
\phi=f(x/\lambda-t)+Zf(-x/\lambda-t),
\end{equation}
where $f$ is a yet unknown function, $\lambda=\sqrt{v^2-u_0^2}$ is the edge mode velocity in the interacting region, $u_0=u(x=0)$ and 

\begin{equation}
\label{15}
Z=\frac{\sqrt{v+u_0}-\sqrt{v-u_0}}{\sqrt{v+u_0}+\sqrt{v-u_0}}.
\end{equation}
Physically, the solution (\ref{14}) shows incomplete screening of a charge in point $x$ by the edge across the gate. The ratio of the screening and screened charges equals $-Z$.

\begin{figure}
\centering
\includegraphics[width=3in]{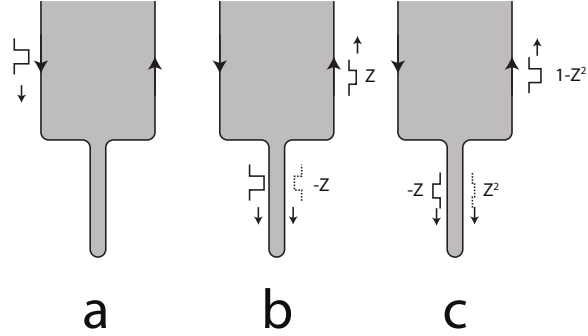}
\caption{a) A charge bump arrives at the interacting region. b) The charge bump on the left is accompanied by the image charge on the right. c) The image charge is reflected at $x=-a$. 
It is screened by another charge bump near $x=a$.}
\end{figure}

The next task is to find $f(z)$ for a given initial condition at $t=-\infty$. The chiral nature of the transport at $|x|>a$ simplifies that task. Indeed, what matters at $t\approx 0$ is only the initial condition $\phi_0(x,t=-\infty)$ at large negative $x$. It will be convenient to us to rewrite $\phi_0(x/v-t)$ as the sum $\phi_0=\sum_{n=-\infty}^{+\infty} g_n(x/v-t)$, where $g_n(z)=\phi_0(z)\theta(n\epsilon-z)\theta(z-[n-1]\epsilon)$ is nonzero only in the interval $(n-1)\epsilon<z<n\epsilon$ for a small 
$\epsilon$. We next solve Eq. (\ref{13}) with the initial condition $\phi(x,t=-\infty)=g_n(x,t=-\infty)$. At $x<-a$ , the solution can be visualized as a bump $g_n(x/v-t)$ that travels towards $x=-a$ (Fig. 2a). Until the bump reaches $x=-a$
at $t=t_n=-a/v-n\epsilon$, the field $\phi(|x|<a,t)$ remains zero. To find $\phi(|x|<a, t>t_n)$ we use the continuity of the electric current $j_i\sim\partial_t\phi_i$ in the point $x=-a$: $\phi(-a-0,t)=\phi(-a+0,t)$. This gives

\begin{equation}
\label{16}
f(x/\lambda-t)=f_0(x/\lambda-t)=\phi_0(x/\lambda-t+a/\lambda-a/v)
\end{equation}
at $t\approx t_n$.
The screening field $Zf(-x/\lambda-t)$ (\ref{14}) appears simultaneously at $x\approx a$. Charge conservation near $x=a$ implies that nonzero $\phi$ appears then at $x>a$ (Fig. 2b). We will not be interested in $x>a$ below.

Eq. (\ref{16}) describes a charge bump\footnote{Strictly speaking, the total charge of the bump is zero, and (\ref{16}) represents an electric dipole.}
 that travels from $-a$ to $a$ with the speed $\lambda$ (Fig. 2b). The screening charge corresponds to a bump that goes with the same speed in the opposite direction. At the time $t=t_n+2a/\lambda$, the screening charge reaches the point $x=-a$. Chirality at $x<-a$ means that the charge cannot penetrate beyond that point. Hence, the charge is reflected. This means that an additional contribution to $f(z)$ appears:

\begin{equation}
\label{17}
f_1(x/\lambda-t)=-Z\phi_0(x/\lambda-t+3a/\lambda-a/v).
\end{equation}
This is accompanied by a new bump of the screening charge $Zf_1(-a/\lambda-t)$, Fig. 2c. That screening charge reaches $x=-a$ at $t=t_n+4a/\lambda$. The same considerations as above show the third contribution to $f$

\begin{equation}
\label{18}
f_2(x/\lambda-t)=Z^2\phi_0(x/\lambda-t+5a/\lambda-a/v).
\end{equation}
Reflection processes continue indefinitely. Such charge bouncing resembles the behavior of wave packets in spin chains from Refs. \cite{spin1,spin2}.
Bringing all contributions to $f$ together, we write

\begin{equation}
\label{19}
f(x/\lambda-t)=\sum_s f_s=\sum_{s=0}^\infty (-Z)^s \phi_0(x/\lambda-t+[2s+1]a/\lambda-a/v).
\end{equation}
Equation (\ref{14}) finally yields at $x=0$

\begin{equation} 
\label{20}
\phi(x=0,t)=(1+Z)\sum_{s=0}^\infty (-Z)^s\phi_0([2s+1]a/\lambda-a/v-t).
\end{equation}

At this point we are ready to find the desired correlation function $G(t)=\langle \phi(0,t)\phi(0,0)- \phi^2(0,0)\rangle$. We use the correlation function of the uniform problem without interaction across the gates

\begin{eqnarray}
\label{21}
G_0(x,t)=\langle\phi_0(x,t)\phi_0(0,0)-\phi_0^2(0,0)\rangle=\ln\frac{\pi T \tau_c}{\sin\left(\pi T[\tau_c+i(t-x/v)]\right)}, \\
\label{22}
G_0(x,t)=-\ln[\frac{\tau_c+i(t-x/v)}{\tau_c}]~~{\rm at}~~T\rightarrow 0,
\end{eqnarray}
where $\tau_c$ is an ultraviolet cutoff scale. We get

\begin{equation}
\label{23}
G(0,t)=\frac{1+Z}{1-Z}\sum_{q=-\infty}^{+\infty}(-Z)^{|q|}[G_0({2qav}/{\lambda},t)-G_0(2qav/\lambda,0)].
\end{equation}

\subsubsection{The case of Fig. 1b} In this case $u(x)=u_0>0$ at $|x|>a$ and is $0$ otherwise. We will reduce that case to Fig. 1a. First, we change the variables in the action:

\begin{equation}
\label{24}
\phi(x,t)=\sqrt{\frac{v+\lambda}{2\lambda}}[\theta(x,t)+Z\theta(-x,t)].
\end{equation}

The Lagrangian density $L_u$ becomes

\begin{eqnarray}
\label{25}
L=-\frac{1}{4\pi}[\partial_t\theta(x,t)\partial_x\theta(x,t)+\lambda (\partial_x\theta(x,t))^2]~~{\rm at}~~|x|>a, \\
\label{26}
L=-\frac{1}{4\pi}\left[ 
\partial_t\theta(x,t)\partial_x\theta(x,t)+\frac{v^2}{\lambda}(\partial_x\theta(x,t))^2-\frac{vu_0}{\lambda}\partial_x\theta(x,t)\partial_y\theta(y,t)\big|_{y=-x}
\right]~~{\rm at}~~|x|<a.
\end{eqnarray}
Next, we rescale the coordinate $x_{\rm old}\rightarrow x_{\rm new}v^2/\lambda^2$ for $|x|<a$. The action (\ref{25},\ref{26}) assumes the form (\ref{3}) with $u(x)=0$ at $|x|>a$,
$u(x)\ne 0$ at $|x|<a$,  $a\rightarrow a\lambda^2/v^2$, $v\rightarrow\lambda$ and the interedge interaction at small $x$ becoming $-\lambda u_0/v$ in place of $u_0$ in the Fig. 1a setup.
Note that the effective interaction  $-\lambda u_0/v<0$ is attractive.
After these changes are plugged into Eq. (\ref{23}), we get

\begin{equation}
\label{27}
G(0,t)=\sum_{q=-\infty}^{+\infty}Z^{|q|}[G_0(2qa,t)-G_0(2qa,0)].
\end{equation}

\section{The speed of the  edge mode}

The shape of the $I-V$ curve depends on three parameters: the distance $a$, the interaction $u_0$ and the speed $v$. The distance $a$ should be known from the device specifications. Thus, we are left with two fitting parameters. 
In fact, after looking at the asymptotic behavior of the current at high or low voltages or temperatures, one is left with a single fitting parameter only. 

To understand how this happens, let us find the current in the limits of high and low voltages and temperatures. A detailed calculation for the geometry of Fig. 1a is given in the Appendix. The results are equations  (\ref{A3},\ref{A4},\ref{A6},\ref{A7}). The Appendix also contains a discussion of the applicability of the perturbation theory, used in the previous section. The conditions on the tunneling amplitude $\Gamma$ are stated in Eqs. (\ref{A8},\ref{A9}). Below we follow a different approach and estimate the current from a renormalization group procedure in the spirit of Ref. \cite{kanefisher}. In that procedure we integrate out fast modes in the quadratic part of the action $\int dt dx (L_{u}+L_d)$ and observe what effect the removal of fast degrees of freedom has on the small tunneling contribution to the action $-\int dt \hat T$. The procedure stops when the energy scale reaches ${\rm max}(e^*V,T)$.

Assume first that $e^*V, T\ll \hbar\lambda/a$. In this case, it is convenient to separate the renormalization group procedure into two stages:

1) The cutoff energy $E_{\rm max}> \hbar\lambda/a\sim hv/a$.

2) The cutoff energy $E_{\rm max}<\hbar\lambda/a\sim hv/a$.\\
\noindent
At the end of the first stage the tunneling amplitude assumes a renormalized value, independent of $T$ and $V$. At the end of the second stage, the renormalized $\Gamma$ is multiplied by an additional factor that depends on the geometry.
In the geometry of Fig. 1a, the factor is $[{\rm max}(e^*V,T)a/\hbar\lambda]^{\nu-1}$. In the geometry of Fig. 1b, the factor is $[{\rm max}(e^*V,T)a/\hbar\lambda]^{\nu\frac{1+Z}{1-Z}-1}$.
Thus, at zero $T$ we obtain

\begin{equation}
\label{28}
{\rm Fig.~1a}:~ I\sim V|\Gamma_{\rm renormalized}|^2\sim V^{2\nu-1};
\end{equation}
\begin{equation}
\label{29}
{\rm Fig.~1b}:~ I\sim V|\Gamma_{\rm renormalized}|^2\sim V^{2\nu\frac{1+Z}{1-Z}-1}.
\end{equation}
At a finite temperature we find the scaling of the zero-bias conductance $G\sim|\Gamma_{\rm renormalized}|^2$

\begin{equation}
\label{30}
{\rm Fig.~1a}:~ G\sim T^{2\nu-2};
\end{equation}
\begin{equation}
\label{31}
{\rm Fig.~1b}:~ G\sim T^{2\nu\frac{1+Z}{1-Z}-2}.
\end{equation}

Now consider the case of ${\rm max}(e^*V,T)\gg \hbar a/\lambda$. In this limit the details of the system beyond the distance $\hbar \lambda/{\rm max}(e^*V,T)$ from the QPC do not matter. At low $T$ one finds

\begin{equation}
\label{32}
{\rm Fig.~1a}:~  I\sim V|\Gamma_{\rm renormalized}|^2\sim V^{2\nu\frac{1+Z}{1-Z}-1};
\end{equation}
\begin{equation}
\label{33}
{\rm Fig.~1b}:~ I\sim V^{2\nu-1}.
\end{equation}
The zero bias conductance scales as

\begin{equation}
\label{34}
{\rm Fig.~1a}:~ G\sim T^{2\nu\frac{1+Z}{1-Z}-2};
\end{equation}
\begin{equation}
\label{35}
{\rm Fig.~1b}:~ G\sim T^{2\nu-2}.
\end{equation}

\begin{figure}
\centering
\includegraphics[width=5in]{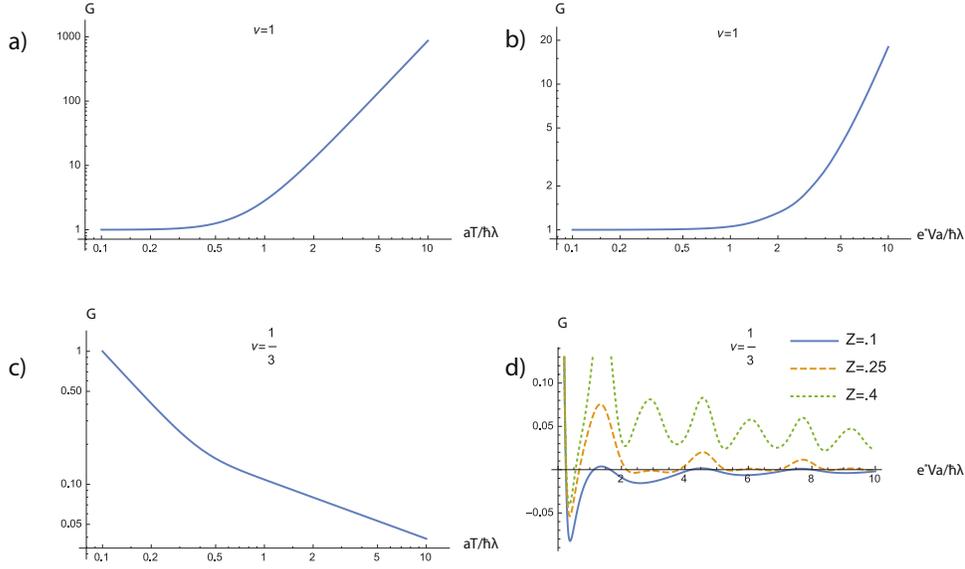}
\caption{Transport in the geometry of Fig. 1a. The voltage is shown in units of $\hbar\lambda/a e^{*}$
and the temperature is shown in units of $\hbar\lambda/a$. The conductance is shown in arbitrary units. 
$T=0.1\hbar\lambda/a$ in panels b) and d). $Z=.4$ in a)-c). A logarithmic scale is used in plots a)-c). a) Zero-bias conductance at $\nu=1$. b) Differential conductance at $\nu=1$. c) Zero-bias conductance at $\nu=1/3$.
d) Differential conductance at $\nu=1/3$.}
\end{figure}

\begin{figure}
\centering
\includegraphics[width=5in]{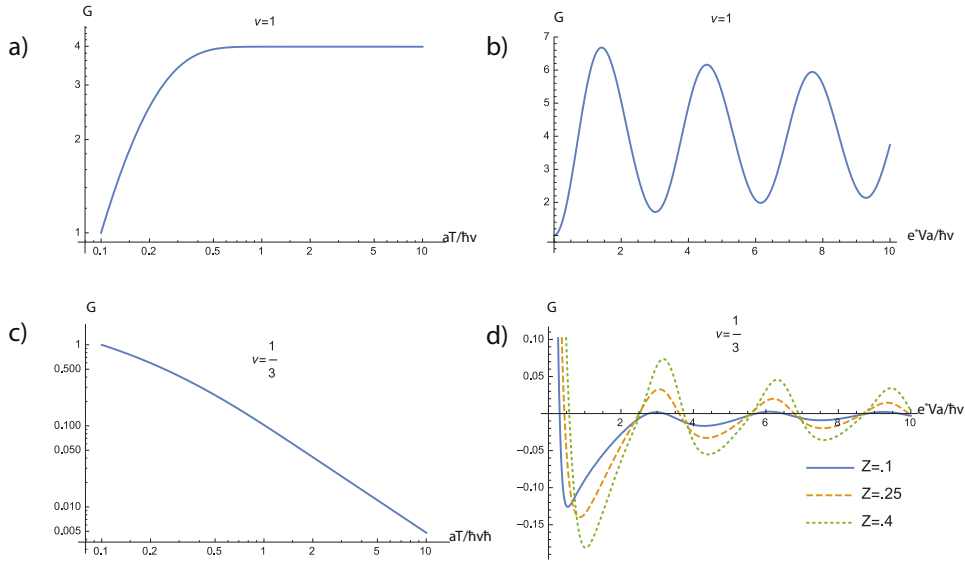}
\caption{Transport in the geometry of Fig.
1b. The voltage is shown in units of $\hbar v/a e^{*}$
and the temperature is shown in units of $\hbar v/a$. The conductance is shown in arbitrary units. 
$T=0.1\hbar v/a$ in panels b) and d). $Z=.4$ in a)-c). A logarithmic scale is used in plots a) and c). a) Zero-bias conductance at $\nu=1$. b) Differential conductance at $\nu=1$. c) Zero-bias conductance at $\nu=1/3$.
d) Differential conductance at $\nu=1/3$.}
\end{figure}

Equations (\ref{29},\ref{31},\ref{32},\ref{34}) can be used to determine $Z$ (\ref{15}) and hence find $u_0/v$. This leaves a single unknown: the edge mode velocity $\lambda$ (or, equivalently, $v$). It can be found by fitting experimental data with theoretical predictions for the current or conductance. A theoretical calculation involves substituting Green's functions (\ref{23},\ref{27}) into the integral (\ref{9}). The results of such a calculation are illustrated in Figs. 3 and 4. The low-temperature $I-V$ curve (Figs. 3b,d and 4b,d) allows an easy estimate of the velocity. Indeed, the current exhibits periodic features in its voltage dependence, such as oscillations in Figs. 3d and 4b,d, due to the bouncing of the image charge, addressed in Section 2. Technically, the oscillations come from the singularities of the correlation functions (\ref{23},\ref{27}). In the geometry of Fig. 1a the locations of the singularities are 

\begin{equation}
\label{d1}
t=\frac{2na}{\lambda}.
\end{equation}
This translates into the current oscillations with the voltage period 

\begin{equation}
\label{36}
\Delta V\sim\frac{\pi\hbar\lambda}{ae^*}.
\end{equation}
We use the order of magnitude sign $\sim$ because the current is not a strictly periodic function due to a strong voltage dependence of the oscillation amplitudes. 
Note that in the geometry of Fig. 1a the oscillations with the period $\pi\hbar\lambda/ae^*$ are dominant at small $Z\ll 1$. As is clear from Fig. 3d, at greater $Z$ the dominant oscillations occur at the double frequency, i.e., half the period. This corresponds to $n=2$ in Eq. (\ref{d1}). 
The frequency change can be understood by looking at the factors $(-Z)^{|q|}$ in Eq. (\ref{23}). At small $Z$ only $q=0,\pm 1$ matter and this corresponds to $n=q=1$ in Eq. (\ref{d1}). On the other hand, at greater  $Z$, higher values of $n$ in Eq. (\ref{d1}) become important and even $n$ get more important than odd $n$.
Indeed, the exponent of Green's function in Eq. (\ref{9}) diverges only in the points (\ref{d1}) with even $n$. Odd $n$ correspond to zeros of the exponent and are responsible for a subleading contribution to the current.  On the other hand, in the geometry of Fig. 1b the exponent diverges in all singularities  and the frequency doubling does not occur (Fig. 4d).
Note also a similarity with the period doubling observed in an interferometry experiment \cite{moty}.

In the geometry of Fig. 1b the singularities are located at $t=2na/v$. The corresponding period of the voltage oscillations

\begin{equation}
\label{37}
\Delta V\sim\frac{\pi\hbar v}{ae^*}.
\end{equation}
Substituting the distance between subsequent maxima of the $I-V$ curves in Fig. 4d into Eq. (\ref{37}), one can quickly estimate the edge velocity.

The physical origin of the current oscillations is most transparent in the limit of strong interactions, $Z\approx 1$, in the geometry of Fig. 1a. Consider a bump of negative charge that appears between $x=-a$ and $x=a$ due to a tunneling event. It propagates with the speed $\lambda$ towards the point $a$. The bump is accompanied by the screening charge $-Zq$ that moves in the opposite direction. Due to chirality of the transport in the noninteracting region $|x|>a$, the screening charge cannot penetrate beyond the point $x=-a$ and is reflected. It reflects and propagates towards the point $a$. The original bump $q$ is partially transmitted through the point $a$ but most of the charge in the amount of $Z^2q$ is reflected and goes towards the point $-a$. At that point the charge bump $Z^2q$ is reflected again. The system behaves in an approximately periodic way: after the charge $q$ arrives to the point $a$ it takes the time $\tau=2\times 2a/\lambda$ for the charge $Z^2 q\approx q$ to reach the point $a$ the next time. Recall that the edge action is quadratic, that is,  the system is an ensemble of harmonic oscillators. Hence, such closed orbits with the period $\tau$ correspond to quasistationary levels of the energies $E_q=2\pi (q+1/2) \hbar/\tau$ with an integer $q$.
Transmission resonances at the QPC occur at $e^*V=E_q-E_0$ and result in periodic features in the $I-V$ curve. The period $\pi\hbar\lambda/2ae^*$ is one half of the right hand side of Eq. (\ref{36}). Equating the above period with the distance between the second and third maxima of the $I-V$ curve with $Z=.4$ in Fig. 3d gives the correct value of $\lambda$ with the accuracy of 8\%. The distance between the third and fourth maxima of the same curve 
gives 5\% accuracy.

\section{Conclusions}

Our approach allows not only a measurement of the edge velocity but also the interaction $u_0$ across the gate. The interaction describes the mutual capacitance of the edge channels on the opposite sides of the gate and depends on the edge geometry. The relevant geometric information includes the width of the gates, the width of the edge channels and their distance from the gates. One can get insight into that information from knowledge of $u_0$. Thus, the proposed experiment can shed light not only on the speed of charge propagation but also on the structure of the edge, including its dependence on the width of the gate and the gate voltage.

\begin{figure}
\centering
\includegraphics[width=1.5in]{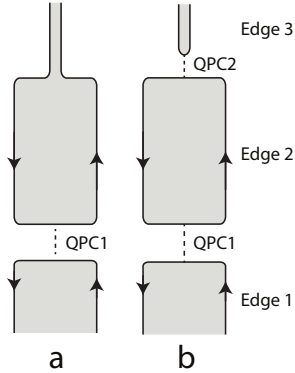}
\caption{a) The same as Fig. 1 but the width of the lower gate does not depend on the distance from the QPC. b) The island of length $a$ is connected to two QHE edges by two quantum point contacts QPC1 and QPC2.}
\end{figure}

Our calculations can be extended beyond the simplest filling factors $1$ and $1/3$. It is also easy to address other related geometries. For example, one can consider the geometry of Fig. 5a, where one of the gates has the same width at all distances from the QPC. The $I-V$ curve is described by the same Eq. (\ref{9}) as in the geometries of Fig. 1 with one modification: $2\nu\langle\phi_u(x=0,t)\phi_u(0,0)-\phi_u^2(0,0)\rangle$ in the exponent is substituted by $\nu[G(0,t)+G_0(0,t)]$, where $G(0,t)$ is given by Eq. (\ref{27}) and $G_0(0,t)$ is given by Eq. (\ref{21}). The same expression describes the $I-V$ curve in the geometry with a straight edge instead of the lower $\Pi$-shaped edge channel.

We assume a constant width of the narrow part of the gates in Fig. 1a. In a realistic sample, the edge width may deviate from the specifications and the width of the narrow section might depend on the distance from the tunneling contact. Our results remain unaffected as long as the width fluctuations are much smaller than the average width of the narrow part of the gate. A significant coordinate dependence of the width may completely change the transport behavior. For example, Fig. 6 illustrates gates whose width depends linearly on the distance from the constriction. The tunneling current is determined by the typical Coulomb interaction within the distance $l_T\sim \hbar \lambda/T$ from the constriction. That interaction is a function of the average width of the gates within the distance $l_T$ from the QPC. Since such a width depends on the temperature, the high-temperature asymptotic of the conductance is not given by a power law.

\begin{figure}
\centering
\includegraphics[width=1in]{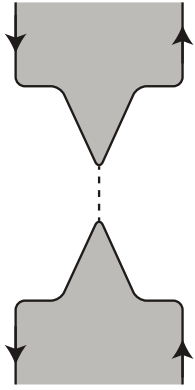}
\caption{The same as Fig. 1a but the width of the narrow sections of the gates depends on the distance from the QPC.}
\end{figure}

We treat tunneling at the QPC in Figs. 1a and 1b as if it occurred at a single point. The validity of such an assumption crucially depends on the total width of the gates being less than $\hbar v/T$ and $\hbar v/eV$. If the width of the gates exceeds $\hbar v/T$ in the geometry of Fig. 1b then the tunneling current is determined solely by the two horizontal portions of the edge immediately above and below the dashed line in Fig. 1b. The system forms an interferometer and the current may oscillate as a function of the voltage bias. In all our calculations we assume that we are away from such a regime.

The discussion at the end of the previous section reveals a connection of the proposed experiment with Coulomb blockade \cite{cb1}-\cite{cb7},\cite{bishara09,fp331}. Indeed, one can use the geometry of Fig. 5b to extract the edge velocities from the locations of the resonant transmission peaks. In contrast to Fig. 5a, there are three QHE edges with two tunneling contacts between them. Each of the edges is defined by a separate gate which can be maintained at a different gate voltage. Thus, one can trace the dependence of the zero-bias conductance on the gate voltage $V_g$ at gate 2 in Fig. 5b. The energy of edge 2 depends on  its charge $e^*N$: 

\begin{equation}
\label{38}
E(N,V_g)=\frac{\pi \nu \hbar v N^2}{2a}-V_ge^* N=\frac{\pi \nu \hbar v (N - e^* V_g  a/\pi\nu \hbar v)^2}{2a}+{\rm const},
\end{equation}
where $v$ is the speed of the edge mode.
At low temperatures the edge is in its ground state with $N={\rm the~nearest~integer~to}~[e^* V_g e a/\pi\nu \hbar v]$. Transmission peaks occur when $E(N,V_g)$ and $E(N+1,V_g)$ are degenerate, i.e., at
$V_g=\pi\nu \hbar v(N+1/2)/e^* a$. This allows finding $v$. The geometry of Fig. 5b is related to that of Fig. 5a though more complicated because of an additional quantum point contact QPC2. One could fabricate a tunable structure that can be continuously changed from the geometry of Fig. 5b into that of Fig. 5a by changing the voltages at the gates which define QPC2. It would be interesting to compare the results for the edge velocity in different regimes in such a structure. Note that the edge structure and hence the velocity $v$ depend on the gate voltage $V_g$ which changes in the Coulomb blockade experiment. All gate voltages are fixed in the experiment with a single QPC.

In conclusion, we propose a method to find the edge mode velocity in the QHE from dc transport through a single QPC. Such an experiment also sheds light on the geometry of QHE edges.

\ack

We acknowledge support by the NSF under Grant No. DMR-1205715.

\appendix

\section{High and low voltages and temperatures}

In this Appendix we compute the current in several limiting cases. In contrast to the main text, we do not omit $\hbar$ from the equations.

The tunneling current in the geometry of Fig. 1a can be obtained by plugging the correlation function (\ref{23}) into Eq. (\ref{9}). The result is rather unwieldy:

\begin{eqnarray}
\label{A1}
I=\frac{e^*|\Gamma|^2}{\hbar^2}\int_{-\infty}^\infty dt (e^{ie^*Vt/\hbar}-e^{-ie^*Vt/\hbar})\left[\frac{\pi T\tau_c/\hbar}{\sin(\frac{\pi T }{\hbar}(\tau_c+i t))}\right]^{2\nu\frac{1+Z}{1-Z}}\nonumber\\
\times\prod_{n=\pm 1,\pm 2,\dots}\left[ \frac{-\sin(2 i n a\pi T/\hbar\lambda )}{\sin(\frac{\pi T}{\hbar} [\tau_c+i(t- 2 n a /\lambda) ])}\right]^{2\nu(-Z)^{|n|}\frac{1+Z}{1-Z}}.
\end{eqnarray}
  This expression greatly simplifies in the limits of high and low voltages and temperatures.

\subsection{$e^*V, T\ll \hbar\lambda/a$.}

This case is easy. The integral is determined by the region $t\sim{\rm min}(\hbar/T, \hbar/e^*V)\gg a/\lambda$. Thus, we can neglect the contributions $2na \pi T/\hbar \lambda$ in the denominator of the integrand. This leaves us with a much simpler integral:

\begin{equation}
\label{A2}
I\sim \frac{e^*|\Gamma|^2}{\hbar^2}\left(\frac{\lambda\tau_c}{a}\right)^{2\nu\frac{1+Z}{1-Z}}\int_{-\infty}^\infty dt (e^{ie^*Vt/\hbar}-e^{-ie^*Vt/\hbar})\left[\frac{a T }{\hbar \lambda\sin(\frac{\pi T}{\hbar}( \tau_c+i t))}\right]^{2\nu},
\end{equation}
where we omit a constant factor of the order of one. At $V\rightarrow 0$, the linear conductance scales as

\begin{equation}
\label{A3}
G\sim \frac{|e^*\Gamma|^2}{\hbar} T^{2\nu-2}\left (\frac{a}{\hbar\lambda}\right )^{2\nu}\left ( \frac{\lambda\tau_c}{a}\right )^{2\nu\frac{1+Z}{1-Z}}.
\end{equation}
At $T\rightarrow 0$, the current behaves as

\begin{equation}
\label{A4}
I\sim e^*\frac{|\Gamma|^2}{\hbar}(e^* V)^{2\nu-1}\left (\frac{a}{\hbar\lambda}\right )^{2\nu}\left ( \frac{\lambda\tau_c}{a}\right )^{2\nu\frac{1+Z}{1-Z}}.
\end{equation}

\subsection{$e^*V$ or $T$~$\gg \hbar \lambda/a$.} 

We start with the case of large $V$. It turns out that the infinite product in Eq. (\ref{A1}) is simply 1. The value of the integral is determined by $t\sim \hbar/e^*V\ll a/\lambda$. If $\pi T /e^* V$ is small then we can neglect the terms $\pi T t/\hbar$ in the sine functions in the infinite product. This shows that the product equals 1 indeed. Alternatively, if $T$ is large enough that $Tt/\hbar>1$ then $Ta/\hbar \lambda\gg 1$ and sine functions in the infinite product can be approximated by exponential functions:

\begin{equation}
\label{A5}
\frac{-\sin(2 i n a\pi T/\hbar\lambda )}{\sin(\frac{\pi T}{\hbar} [\tau_c+i (t- 2 n a /\lambda)])}=\frac{\exp(2|n|a\pi T/\hbar \lambda)}{\exp(2|n|a\pi T/\hbar \lambda-{\rm sign}(n)\pi T t/\hbar)}=\exp[{\rm sign} (n)\pi T t/\hbar].
\end{equation}
Substituting Eq. (\ref{A5}) into Eq. (\ref{A1}) we find that the infinite product reduces to 1 and

\begin{equation}
\label{A6}
I\sim e^*\frac{\tau_c|\Gamma|^2}{\hbar^2}\left (\frac{e^*V \tau_c}{\hbar} \right )^{2\nu\frac{1+Z}{1-Z}-1}.
\end{equation}

Finally, consider $T\gg \hbar \lambda/a$. The same arguments as above show that the infinite product is close to 1 unless $t$ is close to a multiple of the bouncing time $2na/\lambda$, $n\ne 0$. One easily verifies that the contributions to the integral (\ref{1}) from $t\approx 2na/\lambda$, $n\ne 0$ are suppressed by exponentially small factors $\sim\exp(-{\rm const~} aT/\hbar \lambda)$. Thus, we are allowed to substitute the infinite product with 1 and obtain
the zero bias conductance

\begin{equation}
\label{A7}
G\sim \frac{|e^*\tau_c\Gamma|^2}{\hbar^3}\left (\frac{T\tau_c}{\hbar} \right )^{2\nu\frac{1+Z}{1-Z}-2}.
\end{equation}

\subsection{Validity of the perturbation theory.}

We have expanded the current and conductance to the lowest order in the tunneling amplitude $\Gamma$. When is this justified? The perturbation theory works as long as the tunneling current is much smaller than the current $ \nu e^2V/h$
that flows along the edges. The comparison with Eqs. (\ref{A3},\ref{A4},\ref{A6},\ref{A7}) gives us the following inequalities for $\Gamma$:

\begin{eqnarray}
\label{A8}
\Gamma\ll[{\rm max}(T,e^*V)]^{1-\nu}\left(\frac{\hbar\lambda}{a}\right)^\nu \left(\frac{a}{\lambda\tau_c}\right)^{\nu\frac{1+Z}{1-Z}} ~{\rm at}~e^*V,T\ll \hbar\lambda/a\\
\label{A9}
\Gamma\ll [{\rm max}(T,e^*V)]^{1-\nu\frac{1+Z}{1-Z}}\left(\frac{\hbar}{\tau_c}\right)^{\nu\frac{1+Z}{1-Z}}~{\rm at}~e^*V~{\rm or}~T\gg \hbar\lambda/a
\end{eqnarray}

\end{document}